\documentstyle[11pt]{article}

\newcommand{\bez}{\begin{eqnarray*}}
\newcommand{\eez}{\end{eqnarray*}}
\newcommand{\be}{\begin{equation}}
\newcommand{\ee}{\end{equation}}
\newcommand{\beq}{\begin{eqnarray}}
\newcommand{\eeq}{\end{eqnarray}}
\newcommand{\bc}{\begin{center}}
\newcommand{\ec}{\end{center}}

\newbox\grsign \setbox\grsign=\hbox{$>$} \newdimen\grdimen \grdimen=\ht\grsign
\newbox\simlessbox \newbox\simgreatbox \newbox\simpropbox
\setbox\simgreatbox=\hbox{\raise.5ex\hbox{$>$}\llap
     {\lower.5ex\hbox{$\sim$}}}\ht1=\grdimen\dp1=0pt
\setbox\simlessbox=\hbox{\raise.5ex\hbox{$<$}\llap
     {\lower.5ex\hbox{$\sim$}}}\ht2=\grdimen\dp2=0pt
\setbox\simpropbox=\hbox{\raise.5ex\hbox{$\propto$}\llap
     {\lower.5ex\hbox{$\sim$}}}\ht2=\grdimen\dp2=0pt
\def\simgt{\mathrel{\copy\simgreatbox}}
\def\simlt{\mathrel{\copy\simlessbox}}

\title{\vspace{4cm}\large\bf
Cosmic vacuum and large-scale perturbations \\
in the concordant model
}
\author{Arthur D.~Chernin$^{a,b,c}$\\
$^a$Sternberg Astronomical Institute, \\ Moscow University, Moscow, 119899,
Russia, \\
$^b$Tuorla Observatory, University of Turku, FIN-21500,  Finland, \\
$^c$Astronomy Division, University of Oulu, FIN-90014, Finland
}

\date{~}

\begin{document}

\maketitle

\begin{abstract}

\noindent
{\bf
The evolution of cosmological large-scale perturbations is described
in terms of the concordant model based on the recent discovery of
cosmic vacuum. It is demonstrated that the process is robustly controlled
by a few epoch-independent physical parameters. The parameters
guarantee that the initially weak perturbations become nonlinear before
(but near) the termination of gravitational instability by cosmic vacuum.
No fine tuning for the perturbation amplitude is needed.
}
\end{abstract}



\newpage

\section{Introduction}

The existence of cosmic vacuum
(Riess et al., 1998, Perlmutter et al., 1999) implies
(Peebles, 1980,
Zeldovich \& Novikov, 1982, Heath, 1977, Carroll \& Press, 1992) that the
gravitational instability and the
linear growth of large-scale perturbations in matter
are effectively
terminated by the antigravity of cosmic vacuum
at the redshifts $z \le 0.7-0.3$ when
the cosmological dynamics  becomes vacuum
dominated.
The perturbations must obviously become
nonlinear before that moment. On the other
hand, the nonlinear  evolution starts not earlier than at $z \simeq 3-10$
(see again the references above).
This means that the perturbations are finely tuned: their
relative amplitude must reach a unity level within the narrow
redshift interval from $z \simeq 10-3$ to $z \simeq 1$.

The accuracy of such fine-tuning may be seen
from the standard relations for
linear perturbation growth.
The relative amplitude of density perturbation  in cold
dark matter grows as
\begin{equation}
\delta (t) \propto (1 + z)^{-1}, \;\;z < z_{eq},
\end{equation}
\noindent
where $z_{eq}$ is the redshift at the moment when $\rho_R = \rho_D$,
and $\rho_R, \rho_D$ are the densities of radiation and (dark) matter.
For the earlier epoch of radiation domination,
\begin{equation}
\delta (t) \propto  (1 + z)^{-2}, \;\; z > z_{eq}.
\end{equation}
\noindent
To reach a unity level, $\delta  \sim 1$,  within
the redshift interval $10-1$, the amplitude $\delta (t)$ must be tuned
to one part in $10^{16}$
at the epoch of light element production,
to one part in $10^{30}$ at the electroweak energy ($\sim 1$ TeV) epoch,
or to
one part in $10^{60}$ at the Planck epoch.

These fine-tuning considerations (they may easily be refined to take
into account all the details of gravitational instability process on
various scales and at various epochs, etc.) show that the initial cosmological
perturbations that give rise to the formation of the observed large-scale
structure are not completely arbitrary. They are rather
well organized to be in correspondence with the conditions of the whole
cosmic dynamics.
Whenever they are generated, the perturbations `know' from the very
beginning that they produce the structure before (but near)
the termination epoch. The questions are: Why do they know this in advance?
And how exactly are they prepared to yield the condition?

Fine-tuning  in cosmology was first discussed more than
30 years
ago, when Dicke (1970) mentioned that the Universe must
be `extremely finely tuned' to account for the present observed
balance between
the kinetic energy of expansion $K$ and the gravitational potential energy
of cosmic matter $U$. The balance is quantified in terms of
$\Omega$, which is the energy ratio, $\vert U \vert /K$, and also the ratio
of the  total matter density, $\rho$, to the critical density, $\rho_c$.
Thirty years ago,  observational limits on $\Omega$ were described
as $0.1 < \Omega_0 < 10$, and it was argued that this
range implies a very narrow range at earlier
epochs. It is estimated that $\Omega$ departs from unity
by one part in $10^{16}$
at the  epoch of light element production,
or to
one part in $10^{60}$ at the Planck epoch.
Why was there such a remarkably fine balance between the kinetic and
potential energies in the `initial conditions' for the cosmic expansion?
This argument is now known
as the flatness problem, since the energy ratio is associated with the
sign of the spatial curvature in the Friedmann model.

In an analysis of the flatness problem (Chernin, 2003),
one and only one order-of-unity constant parameter is recognized
behind the apparently fine-tuned evolving energy ratio. In this way, the
real meaning of the problem is clarified and the problem is
reduced to the question about the physical nature of the control parameter.
In the present paper, I try a similar approach to the structure formation
problem basing on the Friedmann standard model and the concordant
dataset.

Note that inflation models (see a book by Linde, 1990)
suggest elegant widely accepted solutions to both flatness problem
and perturbation problem with the use of the near-Planckian physics
in the very early universe. These hypothetical pre-Friedmann stages of
cosmic evolution are out of the scope of my discussion here.

\section{Physical parameters of the concordant model}

The current standard cosmology is the Friedmann model together with the
 concordant dataset (see, for instance, a recent review by Peebles
and Ratra, 2002). The observational data on cosmic vacuum (V),
dark (D) matter, baryons (B) and radiation (R) are represented in the
 model by four constant parameters $A_V, A_D, A_B, A_R$ that come
from the Friedmann `thermodynamic' equation:
\begin{equation}
A = [ \kappa \rho a^{3(1+w)}]^{\frac{1}{1 + 3w}}.
\end{equation}
\noindent
Here $w = p/\rho$ is the constant pressure-to-density ratio for a
given cosmic component in the co-moving volume; $w = -1, 0, 0, 1/3$ for
vacuum, dark matter,
baryons and radiation, respectively; $\rho$ is the corresponding
energy density; $a(t)$  is the scale factor
or/and the curvature radius of
the isotropic 3-space; $\kappa = 8\pi G/3 = (8\pi/3)
M_{Pl}^{-2}$; the Planck energy $M_{Pl} = 1,2 \times 10^{19}$ GeV.
The units are used in which the Boltzmann constant $= c = \hbar =1$.

Estimated with the concordant dataset, the four parameters prove to be
numerically identical, on the order of magnitude (Chernin, 1968, 2002):
\begin{equation}
A_V \sim A_D \sim A_B \sim A_R  \sim 10^{60 \pm 1 } M_{Pl}^{-1} .
\end{equation}
\noindent
The identity of the parameters is hardly a simple arithmetical coincidence;
it is rather a remarkable empirical fact in which a kind of
epoch-independent internal symmetry between vacuum
and cosmic matter reveals (Chernin, 2002).

The Friedmann `dynamical' equation that incorporates these parameters,
\begin{equation}
\dot {a}^2 = (A_V/a)^{-2} +  A_D/a + A_B/a + (A_R/a)^2 - k,
\end{equation}
\noindent
contains also the fifth parameter which is the curvature discrete parameter
$k = 1, 0, -1$, for positive, zero and negative 3-space curvature,
correspondingly.
This parameter cannot yet be determined directly from observation, so that
flat and non-flat versions of the model must be
considered, generally. Because of its simplicity, the flat model is usually
used as the preferable one in the structure formation theory.

It is more important that the flat model is a good
approximation as well  for the dynamics and geometry of the non-flat models.
Possible deviations from the flat model at the present epoch are strongly
constrained by the concordant dataset.
As for any other epochs, severe
theoretical bounds on possible deviations
from parabolic dynamics and flat 3-geometry are given by the concordant
model (Chernin, 2003). The bounds come directly from Eqs.(4,5). Indeed,
the maximal possible deviations from the flat model are found to be
near the present epoch; they can be quantified
with the extremal (maximal or minimal) density ratio $\Omega_{ex}$:
\begin{equation}
\Omega_{ex} \simeq [1 - {\frac{1}{2}} k (A_V/A_D)^{2/3}]^{-1}.
\end{equation}

In accordance with the symmetry relation of Eq.(4), $A_V \sim A_D$,
and so  $\vert \Omega - 1 \vert \simlt 1$ for any times.

It is Eq.(6) that enables to recognize the control physical parameter
$A_V/A_D$ which is behind
the phenomenon of flatness or the fine-tuned
energy ratio, as is mentioned in Sec.1.

\section{Linear perturbations}

It was first mentioned by Zeldovich (1965) that
the growth rate of linear perturbations in the flat model could be obtained
from the Friedmann solution with the use of the variation of its two
constants of integrations. The two constants
are the curvature parameter $k$ (or the energy $E = K + U$, in the
Newtonian treatment) and the time moment of zero
radius $t_i$. In the unperturbed model, $k = E = t_i = 0$. If in
a local perturbation volume $k \ne 0, t_i = 0$, the relative density
perturbation increases as in Eqs.(1,2). If in the perturbation
volume $k = 0, t_i \ne 0$, the relative density perturbation decreases
as $\delta \propto t^{-1}$. The shape of the local perturbation
volume does not matter in the linear analysis; as for its size, it
must be larger than the Jeans length in the growing mode.

Having this in mind, let us look more closely at the growing,
most interesting,  mode of perturbations. The relative amplitude of
density perturbation may be expressed via the density ratio $\Omega$:
\begin{equation}
\delta (t) = (\rho - \rho_c)/\rho_c = \Omega - 1.
\end{equation}
\noindent
Here $\rho$ is the density of the perturbation,
\begin{equation}
\kappa \rho = A_V^{-2} + A_D a^{-3} + A_B a^{-3} + A_R^2 a^{-4},
\end{equation}
\noindent
described by the Friedmann equation (5) with $k \ne 0$;
$\rho_c$ is the unperturbed ($k = 0$) density which is also the
`critical' density for the same equation.
The relation of Eq.(7) between the amplitude $\delta$ and density
parameter $\Omega$ provides a clear analogy to the analysis
of Sec.2.

>From Eq.(5), one can find the time evolution of the perturbation
amplitude in the linear approximation:
\begin{equation}
\delta (t) = \Omega - 1 = k (\kappa \rho a^2)^{-1}
[1 - k (\kappa \rho a^2)^{-1}]^{-1},
\end{equation}
\noindent
where the scale factor $a(t)$ is given by Eq.(5) with $k=0$;  actually the
difference between the perturbed and unperturbed scale factors does
affect the right-hand side of Eq.(9), in this approximation; same is
for the density in the right-hand side of Eq.(9).

It is easy to see that the standard relations of Eqs.(1,2) are contained in
Eqs.(8,9) as two limiting cases of matter and radiation domination.
In addition, it is also seen from Eqs.(8,9) that the
perturbation growth is terminated  when the vacuum energy
density becomes larger than the matter densities. Indeed, in the limit
$z <<  z_V = (\rho_V/\rho_D(t_0)^{1/3} -1$ when vacuum dominates completely,
\begin{equation}
\delta (t) \propto a^{-2} \propto \exp (-2t/A_V).
\end{equation}

The extremal value of the amplitude $\delta (t)$ is reached near the
termination at the redshift
\begin{equation}
z_{ex} = 2^{1/3} a(t_0)/(A_V^2 A_D)^{1/3} -1 =
(2 \rho_V/\rho_D)^{1/3} -1.
\end{equation}

The maximal (positive, $k = 1$) and minimal (negative, k = - 1)  density
contrast  is given by the relation
\begin{equation}
\delta_{ex} =  {\frac{1}{2}} k (A_V/A_D)^{2/3}
[1 - {\frac{1}{2}} k (A_V/A_D)^{2/3}]^{-1}.
\end{equation}

The extremal value of the amplitude $\delta$ is completely determined by
two parameters:
the discrete parameter $k$ and the constant ratio $A_V/A_D$, which is of
the order
of unity, according to Eq.(4). As a result, one has:
$\delta_{ex} \simeq 1$ for $k = 1$ and
$\delta_{ex} \simeq -1/3$ for $k = - 1$.
It is easy to see that the density contrast in the dark matter distribution
is not less than the total density contrast, $\delta_D \simgt \delta$, at
this moment.

Since the linear analysis becomes invalid near $z \simeq z_{ex}$, the
numbers for $\delta_{ex}$ should be understood as approximate. But it is
the very fact of nonlinearity that is most important here: the perturbations
do come to the nonlinear regime, and only after that the antigravity of
vacuum wins over the gravity of matter.

It is worthwhile to note that Eqs.(7-12) can be re-written to describe
(with a similar result) the
perturbation evolution in an open ($k = -1$) model. In this case,
perturbations are characterized by $k = 1$ and $k = 0$.

\section{Conclusions}

As we see, the set of Eqs.(7-12) gives a self-consistent description of the
perturbation evolution: the linear perturbations develop in the
standard time rate and reach a unity-level amplitude to the appropriate
epoch. The evolution is
determined and robustly controlled by two constant physical
parameters: these are the curvature discrete parameter $k$ and the
ratio $A_V/A_D$. The both are unity in absolute value or about unity:
$\vert k \vert = 1, A_V/A_D \sim 1$. The first of them quantifies
the perturbations,
the second one comes from the symmetry relation between vacuum and matter.
In combination, they guarantee a quantitative certainty to
the picture of structure formation. At earlier stages of the perturbation
evolution, the constants $A_R$ and $A_B$ are also in the play.
No fine-tuning for the perturbation amplitude is evidently needed, in
this picture at any stage. No specific assumptions about the epoch when
the perturbations are generated are necessary either.

The relations of Sec.3 indicate that
the amplitude of the growing density perturbations
does not depend on their spatial scales; such a feature is known as the
Harrison-Zeldovich spectrum. This (and not only this)
indicates  that the perturbations of Sec.3 are not arbitrary. As
is obvious from Eqs.(7-9), they are prepared in a special way:
each of the perturbation volume is a part of a uniform universe with
the same set of parameters $A_V, A_D, A_B, A_R$ as in the `unperturbed'
universe, but with its own curvature parameter $k$ which is $+ 1$ for
the areas of enhanced density and $- 1$ for the areas of lower density.
The whole distribution of the weak perturbations is represented by a
linear superposition of such patches with different sizes and shapes
over all the cosmic space.

Generally, the superposition may include not only the growing mode, but
also the decreasing mode of
perturbations associated with the parameter $t_i$ which may be a
function of the Lagrangian coordinates $\chi$.
As is well-known, an arbitrary choice of initial
conditions for perturbations assumes the existence of two independent
arbitrary functions of $\chi$, one of which is scalar and the other is
vector,  to describe the density distribution and velocity field,
correspondingly. The choice is essentially
restricted  by the parameters $k$ and $A_V/A_D$, in the description of Sec.3,
while the function  $t_i(\chi)$ remains arbitrary, and its dynamical effect
is rapidly vanishing with time.
The first two parameters determine completely the
fate of the perturbations,
independently of $t_(\chi)$.

This is a possible answer to the question about the structure
of the initial perturbations that is put in Sec.1.

As for the other question of Sec.1 ("Why do they know..."), the
considerations above suggest that there is a
link between the local space-time conditions in the universe and its
global structure and behaviour. The perturbation analysis
reveals only one of the aspects of the link. In this case,
the local-global link is recognized in terms of a few constant physical
parameters that seemingly run the cosmic grand design on its various
space-time scales.

The parameters $k, A_V, A_D, A_B, A_R$ have a phenomenological
status in the concordant model,
and their numerical values are determined (or can be
determined, in principle) from observations of the present-day universe.
But their ultimate  nature is in the fundamental physics that
gave rise to the origin of the Friedmann universe
with its cosmic species -- vacuum, dark matter, baryons, radiation, etc.
This is probably the electroweak-scale physics in
which the symmetry relation of Eq.(4) is argued to have its roots
(Chernin, 2002).

\newpage

\section*{References}

Carroll, S.M., Press, W.H., 1992, Annu. Rev. Atron. Astrophys. 30, 499.

Chernin, A.D., 1968, Nature 220, 250.

Chernin, A.D., 2002, New Astron. 7, 113.

Chernin, A.D., 2003, New Astron. (accepted); preprint: astro-ph/0112158.

Dicke, R.H., 1970, {\em Gravitation and the Universe.} Amer. Phil. Soc.,
Philadelphia.

Heath, D.J., 1977, MNRAS 179, 351.

Linde, A., 1990, {\em Particle physics and inflationary cosmology.}
AIP, New York.

Peebles, P.J.E., 1980, {\em The Large Scale Structure of the Universe},
Princeton Univ. Press, Princeton.

Peebles, P.J.E., Ratra, B., 2002, preprint: astro-ph/0207347.

Perlmutter, S. et al., 1999, ApJ, 517, 565

Riess, A.G. et al., 1998, AJ, 116, 1009.

Zeldovich, Ya.B., 1965, Adv. Astron. Ap. 3, 241.

Zeldovich, Ya.B., Novikov, I.D., 1982. {\em The Structure and Evolution of the
Universe.} (The Univ. Chicago Press, Chicago and London).

\end{document}